\documentclass{blois}
\usepackage{enumerate}
\usepackage{lineno}
\def\Journal#1#2#3#4{{#1} {\bf #2}, #3 (#4)}

\def\NIMA{{\em Nucl. Instrum. Methods} A}

\def\PRL{\em Phys. Rev. Lett.}
\def\PRD{{\em Phys. Rev.} D}

\def\EPJC{{\em EPJ} C}

\def\be{\begin{equation}}
\def\ee{\end{equation}}
\def\bea{\begin{eqnarray}}
\def\eea{\end{eqnarray}}

\def\40K{$^{40}$K}

\newcommand{\Photo}{}

\begin{document}

\vspace*{4cm}
\title{Dark matter search with the SABRE experiment}

\author{ Paolo Montini for the SABRE Collaboration }

\address{Dipartimento di Fisica, Sapienza Universit\`a di Roma and INFN sezione di Roma,\\
Piazzale Aldo Moro 5, 00185 Roma (IT)}

\maketitle\abstracts{
The SABRE (Sodium Iodide with Active Background Rejection) experiment will search for an annually modulating signal from dark matter particles in our Galaxy using an array of ultra-pure NaI(Tl) detectors surrounded by an active scintillator veto to further reduce their intrinsic background. The first phase of the experiment is the SABRE Proof of Principle (PoP), a single 5kg crystal detector operated in a liquid scintillator filled vessel at Laboratori Nazionali del Gran Sasso (LNGS). The PoP installation is underway with the goal of running in 2018 and performing the first in situ measurement of the crystal background and testing the veto efficiency, thus validating the SABRE concept. GEANT4--based Monte Carlo simulations have been developed to estimate the background in the PoP.  The most recent simulation is based on measured or predicted radio-purity levels of the various detector components and includes detailed versions of the detector part geometries. The second phase of SABRE will be twin arrays of NaI(Tl) detectors operating at LNGS and at the Stawell Underground Physics Laboratory (SUPL) in Australia. By locating detectors in both hemispheres, SABRE will minimize seasonal systematic effects. In this work the SABRE PoP activities at LNGS and results from the most recent Monte Carlo simulations will be presented.}

\section{Introduction}
The leading Dark Matter (DM) candidates are Weakly Interactive Massive Particles (WIMPs) uniformly distributed in a halo surrounding our Galaxy. Direct detection experiments aim to detect the low recoil energy ($< 100$ keV) deposited in WIMP--nuclear collisions. The Dark Matter signal in an Earth based experiment is expected to modulate yearly due to the change in the Earth speed relative to the WIMP halo. Results obtained by the DAMA/LIBRA experiment, based on 250 kg of ultrapure NaI(Tl) scintillating crystals, are consistent wit this scenario. They have observed a modulation in the single hit rate with high statistical significance in the energy region 2--6 keV~\cite{dama}.  The interpretation of this result in the WIMP framework is in tension with results obtained by other experiments~\cite{xenon,lux,scdms}. However, these experiments rely on different target materials so a direct comparison of the results is model dependent. A measurement with the same target material is therefore essential to clarify the origin of the modulation. 

\section{The SABRE experiment}
\begin{figure}[htbp!] \centering
 \includegraphics[width=0.32\textwidth]{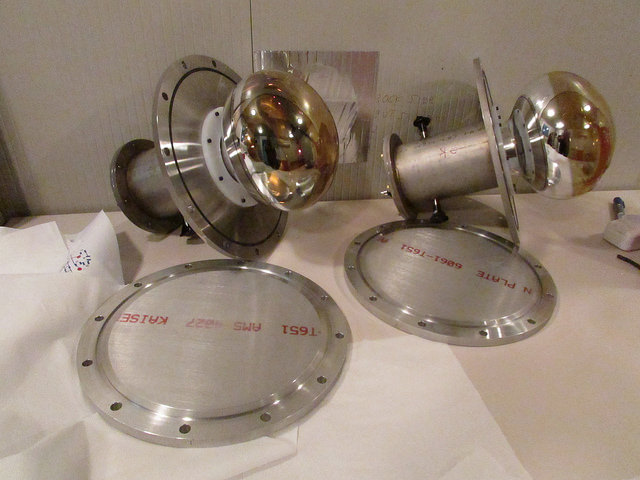}
 \includegraphics[width=0.32\textwidth]{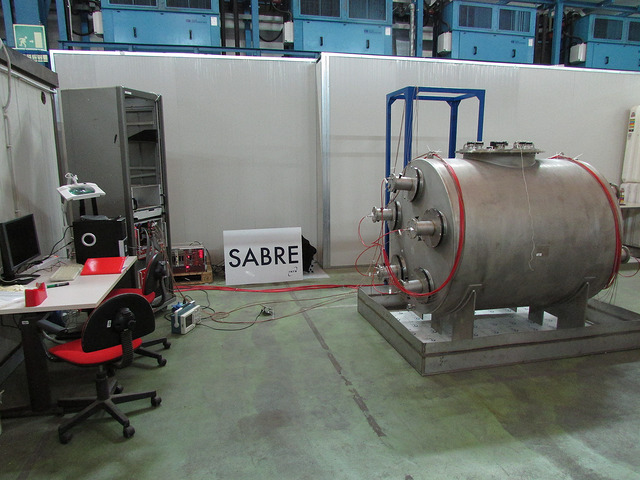}
  \includegraphics[width=0.32\textwidth]{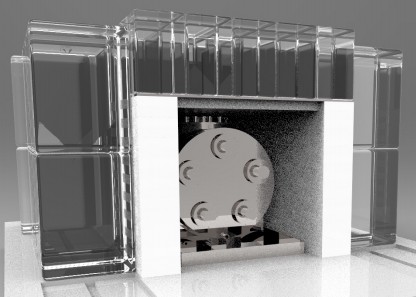}

  \caption{The SABRE PoP setup at LNGS. Veto PMTs mounted on their holders (left). The veto vessel with all the 10 PMTs installed (center). The SABRE PoP shielding design: water tanks (gray) and PE (white) are shown (right).}
\label{pop-lngs}
\end{figure}

The SABRE experiment aims to search for DM through the modulation signal by using ultra--radiopure NaI(Tl) detectors operated within a liquid scintillator active veto.
In order to detect the modulation signal with high sensitivity the SABRE strategy is based on:
\begin{enumerate}[-]
\item background well below the value of 1 cpd/kg/keV achieved by DAMA/LIBRA. For this reason the ultra--radiopure NaI(Tl) crystals will be deployed in a liquid scintillator active veto.
\item low energy threshold achieved by using high QE low--radioactivity PMTs directly coupled to the crystals
\item double location both in Northern and Southern hemispheres in order to identify any possible seasonal effect
\end{enumerate} 
The development of SABRE is organized in two phases. The first one, named SABRE Proof of Principle (PoP) has the goal of validating the whole SABRE strategy and investigating the crystal background and the veto efficiency. The second phase foresees the installation of twin detectors underground at LNGS (Italy) and at SUPL (Stawell Underground Physics Laboratory, Australia).\\
The SABRE PoP detector is currently being deployed at LNGS and will take data in 2018. 
The crystal detector module will consists of a 5 kg ultra-high purity NaI(Tl) crystal wrapped with PTFE and Lumirror and optically coupled to two 3-inch Hamamatsu R11065-20 PMTs. Crystal and PMTs will be packaged in a low radioactivity, air and light tight, high purity copper enclosure. A high purity copper tube will avoid the contatct between the detector module and the liquid scintillator. The volume inside the copper tube will be flushed with high purity N$_2$ gas in order to prevent radon contamination. The veto system consists on a cylindrical stainless steel vessel (1.4 m $\oslash\ \times $ 1.5 m) filled with $\sim 2$ ton of Pseudocumene (PC) + 3 g/l PPO and read--out by 10 8'' wide FoV Hamamatsu R5912 PMTs. A hybrid passive shielding made of water tanks (80 cm top, 1 m sides), polyethilene (10 cm top and bottom, 40 cm sides), steel (2.5 cm top) and lead (15 cm bottom) will reduce the background due to external gamma radiation. The internal volume of the shielding will be flushed with N$_2$ gas in order to remove radon naturally occurring in the laboratory air.
The NaI(Tl) crystals will be grown at RMD (Boston) starting from high purity Astrograde NaI powder provided by Sigma--Aldrich. The contamination of relevant isotopes in Astrograde powder are 9 ppb for K, $<0.2$ ppb for Rb, $<1$ ppt for U, and $<1$ ppt for Th. A 2 kg crystal of nearly the final diameter has been produced in 2015 and the K concentration resulted to be 9 ppb~\cite{davide}. A 5 kg crystal for SABRE PoP is currently in production. The veto vessel with its 10 PMTs is currently located in a temporary area in Hall B at LNGS (see figure \ref{pop-lngs}). The installation of the shielding in the SABRE experimental area in Hall C is ongoing.   

\section{SABRE PoP background}
\begin{figure}[htbp!] \centering
 \includegraphics[width=0.4\textwidth]{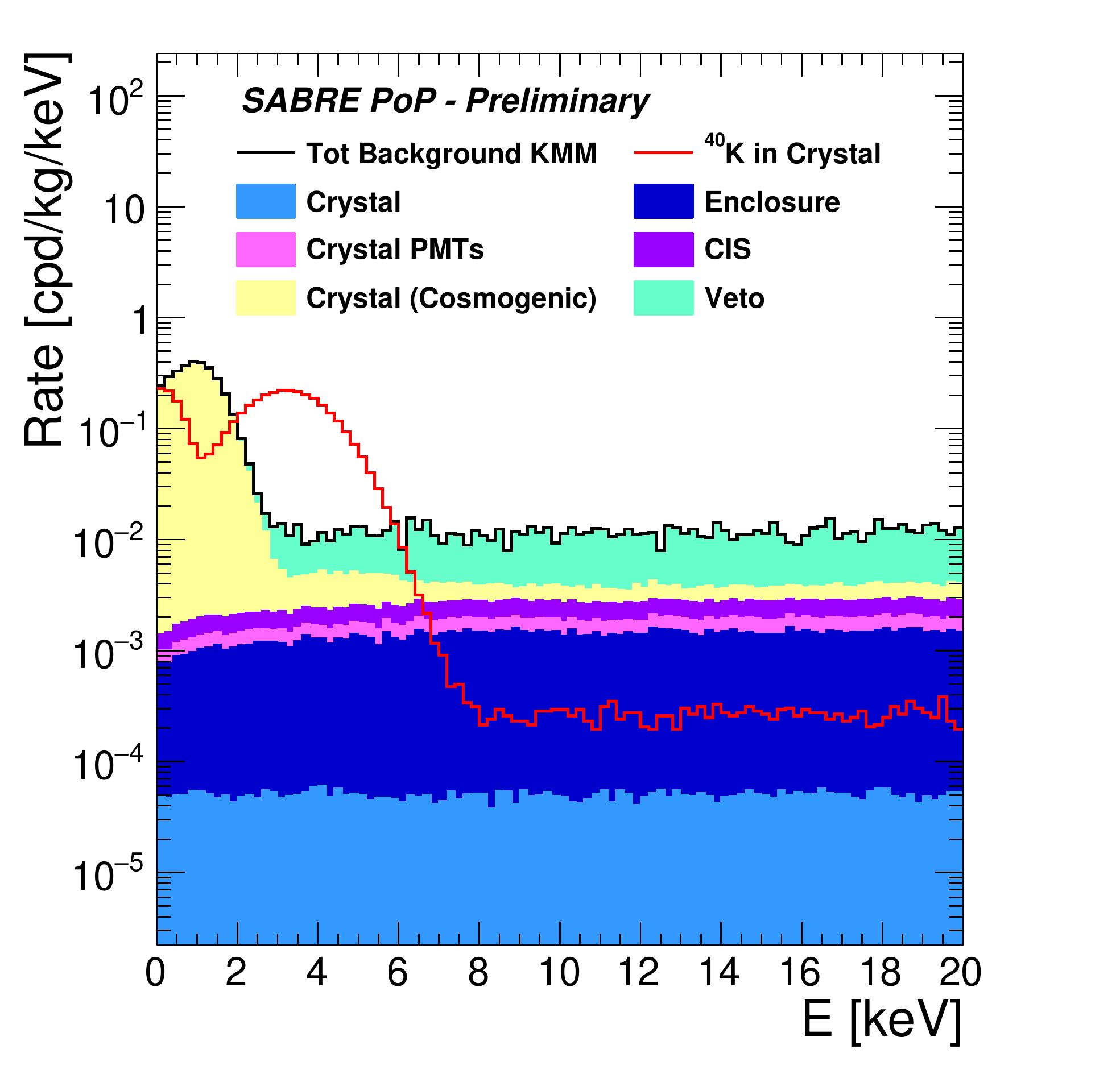}
  \includegraphics[width=0.4\textwidth]{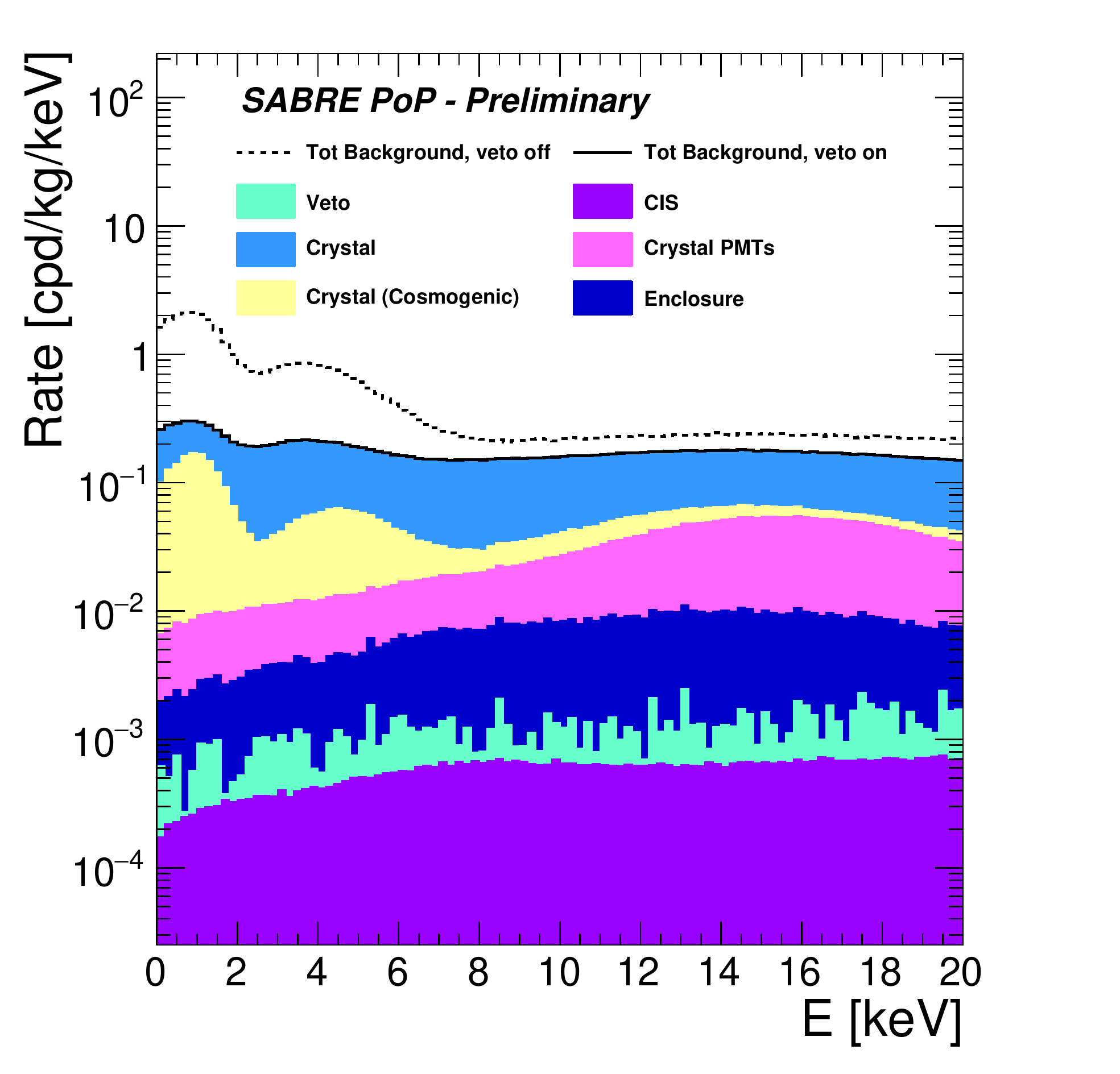}\\
  \caption{Stacked plot of the backgrounds from all SABRE-PoP setup components in a $[0-20]$~keV region in KMM (left) and DMM (right). The red line in KMM represents the signal produced by 10 ppb $^{\mathrm{nat}}$K contamination in the crystal. The dashed black line in DMM indicates the total background with veto off. Cosmogenic activation has been evaluated after 60 (180) days underground for KMM (DMM).}
  \label{fig:background}
\end{figure}

All the backgrounds due to the intrinsic radioactivity and cosmogenic activation of the crystal itself~\cite{davide,damacrys,anais}, the PMTs \cite{xenonpmt} and  the surrounding materials~\cite{cuorecopper} have been simulated by means of a GEANT4--based code. The simulation includes a detailed description of the SABRE PoP detector geometry and materials. All the background due to internal sources has been evaluated up to the external shielding.  \\
A goal of SABRE PoP is to measure the \40K content in the crystals since it is one of the most significant contribution to the background in the energy region of interest. A \40K decay in the crystal gives an energy deposit of 3 keV due to X--ray or Auger electron emission in coincidence with a 1.46 MeV gamma ray which can escape the crystal volume and release its energy in the veto. A potassium--like event in K Measurement Mode (KMM) is defined as an energy deposit between 2 and 4 keV ($1\sigma$ around the 3 keV \40K peak) in the crystal in coincidence with a deposit between 1.28 and 1.64 MeV in the LS ($2.5\sigma$ around the 1.46 MeV \40K peak). The background contributions from all of the SABRE PoP components to the KMM spectrum is reported in a [0--20] keV region in figure~\ref{fig:background} and in table \ref{bulk_total}. Cosmogenic activation is evaluated after 60 days underground. The signal produced by a 10 ppb $^{\mathrm{nat}}$K contamination in the crystal is also shown. \\
A DM--like event (Dark Matter Measurement Mode - DMM) is defined as an energy deposit between 2 and 6 keV in the crystal. In this case the LS is used as an active veto with an energy threshold of 100 keV and $\sim100\%$ light collection efficiency The background in DMM in a [0--20] keV region is also shown in figure~\ref{fig:background} and in table \ref{bulk_total}. 
Crystal intrinsic radioacitivity is the most relevant source of background in DMM, giving $1.5\cdot10^{-1}$ cpd/kg/keV over a total background of $2.0\cdot10^{-1}$ cpd/kg/keV. The veto has an efficiency of $\sim85\%$ for \40K and the total background is reduced by a factor $\sim3.5$. 

\begin{table}[htbp!]
\centering
\label{bulk_total}
\begin{footnotesize}
\begin{tabular}{|c|c|c|c|}
\hline
Isotope & Rate, veto OFF & Rate, veto ON & Rate KMM\\
        & [cpd/kg/keV]   & [cpd/kg/keV] & [cpd/kg/keV] \\
\hline

  Veto                  &  $3.0 \cdot 10^{-2}$ &    $5.7 \cdot 10^{-4}$ & $6.2 \cdot 10^{-3}$ \\  
  CIS(*)                &  $3.7 \cdot 10^{-3}$ &    $4.6 \cdot 10^{-4}$ & $7.7 \cdot 10^{-4}$ \\
  Crystal               &  $3.5 \cdot 10^{-1}$ &    $1.5 \cdot 10^{-1}$ & $5.1 \cdot 10^{-5}$ \\
  Crystal Cosmogenic(*) &  $3.0 \cdot 10^{-1}$ &    $3.9 \cdot 10^{-2}$&  $1.8 \cdot 10^{-2}$\\
  CrystalPMTs           &  $1.2 \cdot 10^{-2}$ &    $8.1 \cdot 10^{-3}$ & $3.7 \cdot 10^{-4}$\\
  Enclosure(*)          &  $9.5 \cdot 10^{-3}$ &    $3.6 \cdot 10^{-3}$ & $1.3 \cdot 10^{-3}$\\
\hline
  Total                 &  $7.1 \cdot 10^{-1}$ &    $2.0 \cdot 10^{-1}$ & $2.7 \cdot 10^{-2}$\\
  \hline
  $^{40}$K signal in KMM          &  -- &   --  & $1.9 \cdot 10^{-1}$ \\
\hline
\end{tabular}
\end{footnotesize}
  \caption{Background rate in the region of interest $[2-6]$~keV from all the SABRE-PoP setup component with veto off and on respectively. Background in $[2-4]$~keV in KMM is also reported. (*)Cosmogenic backgrounds are computed after 180 days underground (60 days in KMM).}
\end{table}

\section{SABRE expected sensitivity}
\begin{figure}[htbp!] \centering
 \includegraphics[width=0.5\textwidth]{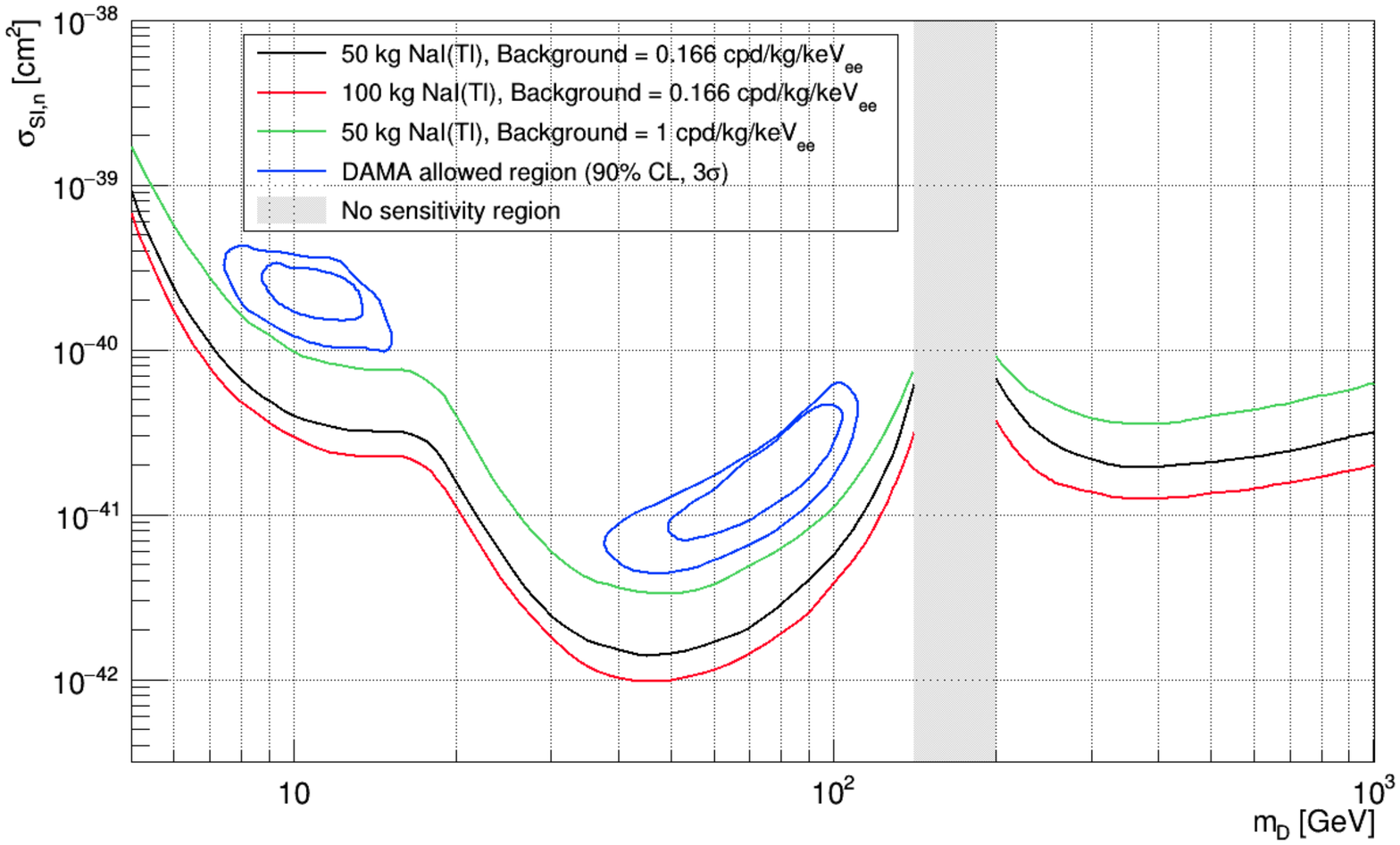}
  \caption{The SABRE expected sensitivity for different detector masses and different background hypothesis. The ``no sensitivity'' region region correspond to the region where the modulation amplitude goes to zero.}
  \label{fig:sens}
\end{figure}

The SABRE sensitivity to the modulation of the WIMP--induced nuclear recoil rate in the 2-6 keV$_{\mathrm{ee}}$ energy range has been evaluated assuming a standard halo model~\cite{freese}. A 90\% C.L. expected upper limits have been evaluated for two different background assumptions (0.2 and 1 cpd/kg/keV) and two different detector masses (50 and 100 kg). The SABRE sensitivity is shown in figure \ref{fig:sens}. SABRE will be sensitive to both the regions extrapolated from a fit of the DAMA/LIBRA data~\cite{savage} in three years of data taking. The plot shows clearly that the suppression of the background has a larger impact on the sensitivity rather than doubling the detector mass. 

\section{Conclusions}
SABRE is an international effort (about 40 physicists from 11 institutions in Australia, Italy, UK and USA) towards the detection of the expected annual modulation of DM particles using an array of ultra--radiopure NaI(Tl) detectors within an active LS veto. The first phase of the project (SABRE PoP) aims to fully characterize the radiopurity of the crystals and the veto efficiency and is currently in the installation phase  at LNGS. The second phase foresees the installation of twin detectors in both the hemispheres (LNGS and SUPL) for a unique investigation into DM modulation. Assuming the background of 0.2 cpd/kg/keV evaluated from Monte Carlo simulations and a detector mass of 50 kg SABRE will provide an independent test of the DAMA/LIBRA result within 3 years.


\section*{References}

\begin{thebibliography}{99}

\bibitem{dama} R. Bernabei {\it et al} [DAMA Collaboration], \Journal{\EPJC}{73}{2648}{2013}.
\bibitem{xenon} E. Aprile {\it et al} [XENON 100 Collaboration], \Journal{\PRL}{109}{181301}{2012}.
\bibitem{lux} D. S. Akerib [LUX Collaboration] {\it et al}, \Journal{\PRL}{112}{031303}{2014}.
\bibitem{scdms} R. Agnese {\it et al} [SuperCDMS Collaboration], \Journal{\PRD}{92}{072003}{2015}.
\bibitem{davide} D. D'Angelo [SABRE Collaboration], \Journal{\em{PoS}}{NOW 2016}{086}{2017}.
\bibitem{damacrys} R. Bernabei {\it et al}, \Journal{\NIMA}{592}{297}{2008}.
\bibitem{anais} J. Amar\'e {\it et al}, \Journal{\EPJC}{76}{439}{2016}.
\bibitem{xenonpmt} E. Aprile {\it et al}, \Journal{\EPJC}{75}{546}{2015}.
\bibitem{cuorecopper} C. Alduino {\it et al}, \Journal{\EPJC}{77}{13}{2017}.
\bibitem{freese} K. Freese {\it et al}, \Journal{\em{RMP}}{85}{1561}{2013}.
\bibitem{savage} C. Savage {\it et al}, \Journal{\em{JCAP}}{0904}{010}{2009}.

\end{thebibliography}
\end{document}